\documentclass[prl,twocolumn,superscriptaddress,preprintnumbers,showpacs]{revtex4-1}

\usepackage{graphicx}
\usepackage{amsmath}
\usepackage{amsfonts}
\usepackage{braket}

\newcommand{\cevns}{CE$\nu$NS}

\begin{document}
\title{First Demonstration of a Scintillating Xenon Bubble Chamber for Detecting Dark Matter and Coherent Elastic Neutrino-Nucleus Scattering}
\date{\today}
\author{D. Baxter}
\affiliation{Department of Physics and Astronomy, Northwestern University, Evanston, Illinois 60208, USA}
\affiliation{Fermi National Accelerator Laboratory, Batavia, Illinois 60510, USA}
\author{C.J. Chen}
\affiliation{Department of Physics and Astronomy, Northwestern University, Evanston, Illinois 60208, USA}
\author{M. Crisler}
\affiliation{Fermi National Accelerator Laboratory, Batavia, Illinois 60510, USA}
\affiliation{Pacific Northwest National Laboratory, Richland, Washington 99354, USA}
\author{T. Cwiok}
\affiliation{Department of Physics and Astronomy, Northwestern University, Evanston, Illinois 60208, USA}
\author{C.E. Dahl}
\email{cdahl@northwestern.edu}
\thanks{ORCID: orcid.org/0000-0003-1637-2346}
\affiliation{Department of Physics and Astronomy, Northwestern University, Evanston, Illinois 60208, USA}
\affiliation{Fermi National Accelerator Laboratory, Batavia, Illinois 60510, USA}
\author{A. Grimsted}
\affiliation{Evanston Township High School, Evanston, Illinois 60201, USA}
\author{J. Gupta}
\affiliation{Department of Physics and Astronomy, Northwestern University, Evanston, Illinois 60208, USA}
\author{M. Jin}
\affiliation{Department of Physics and Astronomy, Northwestern University, Evanston, Illinois 60208, USA}
\author{R. Puig}
\affiliation{Department of Physics and Astronomy, Northwestern University, Evanston, Illinois 60208, USA}
\author{D. Temples}
\affiliation{Department of Physics and Astronomy, Northwestern University, Evanston, Illinois 60208, USA}\author{J. Zhang}
\email{jianjie.zhang@northwestern.edu}
\affiliation{Department of Physics and Astronomy, Northwestern University, Evanston, Illinois 60208, USA}

\preprint{FERMILAB-PUB-17-062-AE-E-PPD} 

\begin{abstract}
A 30-g xenon bubble chamber, operated at Northwestern University in June and November 2016, has for the first time observed simultaneous bubble nucleation and scintillation by nuclear recoils in a superheated liquid.  This chamber is instrumented with a CCD camera for near-IR bubble imaging, a solar-blind photomultiplier tube to detect 175-nm xenon scintillation light, and a piezoelectric acoustic transducer to detect the ultrasonic emission from a growing bubble.  The time of nucleation determined from the acoustic signal is used to correlate specific scintillation pulses with bubble-nucleating events.  We report on data from this chamber for thermodynamic ``Seitz'' thresholds from 4.2~to 15.0~keV.  The observed single- and multiple-bubble rates when exposed to a $^{252}$Cf neutron source indicate that, for an 8.3-keV thermodynamic threshold, the minimum nuclear recoil energy required to nucleate a bubble is $19\pm6$~keV (1$\sigma$ uncertainty).  This is consistent with the observed scintillation spectrum for bubble-nucleating events.  We see no evidence for bubble nucleation by gamma rays at any of the thresholds studied, setting a 90\% C.L. upper limit of $6.3\times10^{-7}$ bubbles per gamma interaction at a 4.2-keV thermodynamic threshold.  This indicates stronger gamma discrimination than in CF$_3$I bubble chambers, supporting the hypothesis that scintillation production suppresses bubble nucleation by electron recoils while nuclear recoils nucleate bubbles as usual.
These measurements establish the noble-liquid bubble chamber as a promising new technology for the detection of weakly interacting massive particle dark matter and coherent elastic neutrino-nucleus scattering.
\end{abstract}

\maketitle

The detection of single nuclear recoils at the keV scale is the core problem in both direct searches for weakly interacting massive particle (WIMP) dark matter \cite{WIMPs} and the detection of neutrinos via coherent elastic neutrino-nucleus scattering (\cevns) \cite{cevns2}. This signal is unique to WIMPs and neutrinos, enabling low-background searches for these extremely rare scattering events via the discrimination of nuclear recoils (signal) from electron recoils (backgrounds).  Easily scalable liquid-based technologies with this capability have proven effective in extending sensitivity to WIMPs \cite{LUX,PICO60-1,PICO2L-2,PICO60-2,PandaX,Xenon100,Darkside}, but the existing techniques are each limited in at least one dimension:  xenon time projection chambers (TPCs) have relatively weak ($10^{-3}$) electron discrimination \cite{LUX-3H} and are susceptible to beta-decay backgrounds; argon-based detectors have much stronger ($10^{-8}$) discrimination at high energies but rapidly lose discrimination for recoil energies below $\sim$45~keV \cite{hughPSD}; and bubble chambers, which have the strongest demonstrated electron-recoil discrimination at $<10^{-10}$, give virtually no event-by-event energy information \cite{PICO2L-Run1} and must address backgrounds both far above and below the keV scale.

The scintillating bubble chamber inherits both the strong electron discrimination of a bubble chamber and the scintillation-based energy reconstruction of a noble liquid.  It can be understood either as a normal bubble chamber with a noble-liquid target and incidental production and detection of scintillation light or as a noble-liquid detector with the usual charge-to-light or pulse-shape discrimination replaced by does-it-make-a-bubble discrimination and the TPC-style position reconstruction replaced by stereoscopic imaging of bubbles.  The technique promises easy scaling and orders-of-magnitude improvement in background discrimination over existing technologies, making it a compelling candidate for future large-scale WIMP and \cevns~searches.
We report results from a 30-g prototype xenon bubble chamber.  To our knowledge this constitutes the first demonstration in any liquid of simultaneous scintillation production and bubble nucleation by nuclear recoils.

The operating principles of the xenon bubble chamber follow closely those of nonscintillating bubble chambers. As described in the Seitz ``hot spike'' model \cite{Seitz}, bubbles are nucleated in the superheated liquid target of the bubble chamber when a particle interaction deposits a minimum amount of heat $E_T$ inside a critical radius $r_c$.  This critical radius describes the smallest vapor bubble that will spontaneously grow in a superheated liquid, and the thermodynamic or ``Seitz'' threshold $E_T$  is the amount of heat needed to create a vapor bubble of the critical radius.  Both $E_T$ and $r_c$ are readily calculated from the vapor pressure, surface tension, and heat of vaporization of the fluid given the pressure and temperature of the superheated liquid \cite{CIRTE}.
In xenon at 30.0~psi absolute (2.07~bar absolute), our two operating temperatures of $-60^\circ$~C and $-55^\circ$~C give $E_T$ of 8.3~and 4.2~keV, with $r_c$ of 32~and 24~nm, respectively \cite{REFPROP}.

In a nonscintillating bubble chamber, the bubble nucleation criteria are met by nuclei above a recoil energy threshold that is typically 1--2 times $E_T$, depending on the target fluid and recoil species \cite{CIRTE, PICO2L-Run1, PICO60-1}.  The difference between the thermodynamic and recoil energy  thresholds may be attributed to energy losses outside the critical radius, due to a combination of recoil range, thermal diffusion \cite{bubblethermalsims}, and radiative losses. Recoiling electrons, which have a much lower stopping power, are inherently unable to create nucleation sites when $E_T$ is greater than a few keV \cite{couppscience}. Nonscintillating bubble chambers have demonstrated bubble-nucleation probabilities for electron recoils as low as $2.2\times10^{-11}$ at $E_T$=3.3~keV in C$_3$F$_8$ \cite{PICO2L-2}, and $5\times10^{-8}$ at $E_T$=7~keV in CF$_3$I \cite{PICO60-1}, with an exponential dependence on $E_T$ in both cases.  The difference in gamma sensitivity for the two fluids is attributed to the iodine in CF$_3$I, due to the potential for cascades of Auger emission from iodine giving a large local energy deposition \cite{AugerBubbles, AlansThesis}.

The expectation for xenon, before considering scintillation, is a gamma sensitivity very similar to CF$_3$I at a given $E_T$.  In an efficient scintillator such as xenon, however, the loss of energy to scintillation light may significantly reduce bubble nucleation.  An early xenon bubble chamber reported no bubble nucleation by gamma rays at thermodynamic thresholds as low as $\sim$1~keV in pure xenon, while the same chamber with 2\% ethylene to quench the production of scintillation light saw bubble tracks as expected at that threshold \cite{Glaser}.  Nuclear recoils, on the other hand, inherently lose most of their energy directly to heat \cite{Lindhard}.  Based on a recent fit of the Lindhard model in xenon \cite{LUX-DD}, a 10-keV xenon recoil loses only 2.1~keV through electronic channels (generating ionization and scintillation) and 7.9~keV through nuclear stopping (i.e., heat).  Scintillation losses should therefore appear as a minor shift in the nuclear recoil bubble nucleation threshold relative to $E_T$ but as a significant decrease in the already very small bubble nucleation probability for electron-recoil events.

We do not expect the superheated state of the liquid to affect scintillation production, as the scintillation time scale [$O(10)$~ns] is much shorter than the bubble growth time scale [$O(1)$~$\mu$s].  
The liquid xenon in our system does have a 10\% lower density than in a typical TPC due to the elevated temperature, which calibrations by the XMASS Collaboration indicate corresponds to a roughly 10\% decrease in the scintillation yield \cite{XMASS}.

\begin{figure}
\begin{center}
\includegraphics[width = 250pt, clip=true]{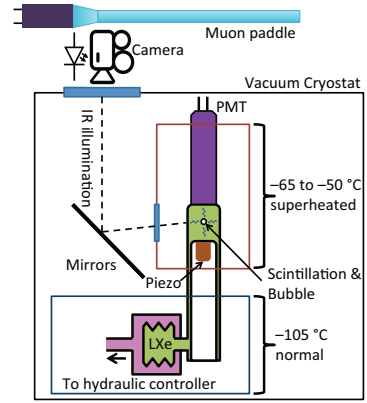}
\end{center}
\caption{\label{F-schematic}Schematic diagram of the 30-g prototype xenon bubble chamber, as described in the text.  The bubble chamber has no buffer fluid and relies on a thermal gradient in the xenon space to achieve superheated xenon in the target region with stable xenon liquid in the plumbing below.}
\end{figure}

A schematic of the experimental setup is shown in Fig.~\ref{F-schematic}.  The target volume is a ``warm'' ($-50^\circ$~C to $-65^\circ$~C) xenon-filled bulb of diameter 24~mm and height 27~mm holding a 30-g xenon target.  The bulb is bounded by two fused-quartz vials, and the xenon extends downward though a 0.5-mm-wide, 80-mm-long annulus between the vials to the ``cold'' ($-105^\circ$~C) volume.  A steep temperature gradient in the middle of the annular section separates the superheated xenon, which sees only fused-quartz surfaces, from the stable liquid in the plumbing below, eliminating the need for a buffer fluid to isolate the superheated target.

Both inner and outer quartz vials are sealed with indium wire to a stainless steel flange, with commercial all-metal seals on the remaining cold plumbing.  Cold components include an absolute pressure transducer, edge-welded bellows for pressure control, and a high-purity cryogenic valve to isolate the xenon space.  Both temperature regions are housed in a vacuum cryostat with an aluminum cold finger to a liquid nitrogen bath, with separate heaters for the two temperature zones maintaining temperatures within $0.1^\circ$~C of their respective set points.  Each temperature region is enclosed in an aluminum radiation shield surrounded by multiple layers of superinsulation, except for a 10-mm-thick heat-sunk sapphire window in the warm radiation shield to allow imaging of the xenon bulb.

\begin{figure*}
\begin{center}
\includegraphics[width = 510pt, clip=true]{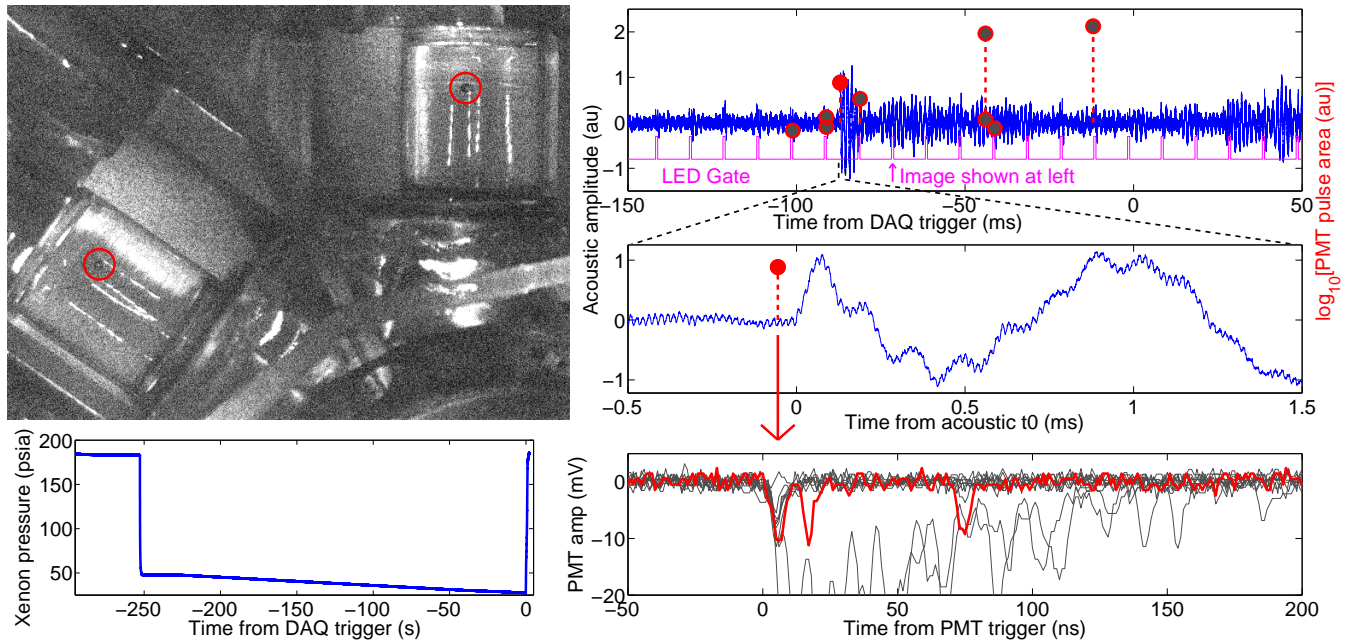}
\end{center}
\caption{\label{F-event}(Sample nuclear recoil event from June 2016.  Clockwise from lower left: (1) Pressure history for the event.  The bubble in this event formed shortly before reaching the target base pressure.  (2) Image of the  xenon target after bubble formation. Two views of the vial are used to reconstruct the 3D position of the bubble. (3) Acoustic record of the event (blue line) along with the camera exposure gate (magenta line).  Xenon PMT triggers appear as red circles, with the $y$~scale indicating the pulse area in log scale for each PMT hit.  (4)  The same as above, zooming in to the time of bubble formation.  (5)  Digitized PMT waveforms.  The red waveform indicates the signal coincident with bubble formation, and the thin gray traces show the waveforms for the other triggers in the top-right plot.  Digitized PMT traces are saved throughout the event, including the time spent compressed prior to expansion.  The $\sim$3-photoelectron pulse in this event is consistent with a low-energy nuclear recoil.}
\end{figure*}

A pair of mirrors inside the cryostat provide stereo views of the target to a CCD camera mounted above the cryostat outside a room-temperature glass view port.  The xenon is illuminated by 955-nm near-IR LEDs flashing in sync with the camera, which takes an 840-$\mu$s exposure every 10~ms.  A solar-blind R6834 Hamamatsu photomultiplier tube (PMT) sits directly above the xenon bulb.  The cap of the outer vial is Corning 7980 UV-grade fused silica to allow the transmission of the 175-nm xenon scintillation light to the PMT.  A piezoelectric acoustic transducer is mounted underneath the inner vial cap, and both the PMT and acoustic transducer are held in direct contact with the quartz vessels.  A 1-cm-thick, 25-cm-wide plastic scintillator paddle mounted above the cryostat provides a rudimentary muon tag,  giving 10\% solid angle coverage directly above the target volume.

The xenon in the target volume cycles between a 200-psi absolute (14-bar absolute) stable liquid state and 30-psi absolute (2-bar absolute) superheated state.  Pressure control is achieved with a hydraulic system using the commercial hydrocarbon blend Dynalene~MV \cite{Dynalene} as the hydraulic fluid, with active feedback from the cold xenon pressure transducer to maintain the xenon pressure within 0.1~psi of the set point.  The pressure cycle for a single event begins in the compressed (stable) state and then expands over a few seconds to 50~psi absolute, corresponding to $E_T$ = 15.9 (6.9)~keV at $-60^\circ$~C ($-55^\circ$~C).  The pressure then ramps down at 0.1--0.5~psi/s to a base pressure of 30~psi absolute, where it remains until a bubble forms.  When a bubble is detected, the chamber rapidly recompresses to the stable state and then sits compressed for 60~s before beginning the cycle again.  The compression is triggered by transients in either the camera images or the pressure sensors in the hydraulic system.

\begin{figure}
\begin{center}
\includegraphics[width=250pt, clip=true]{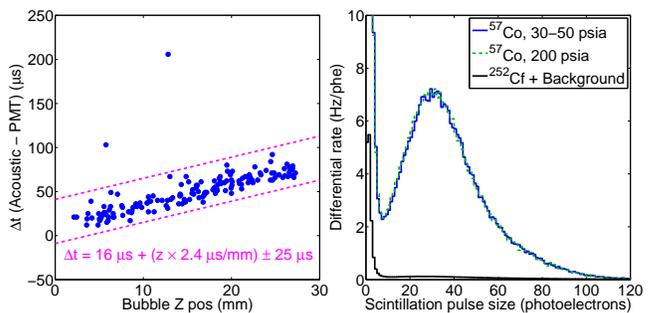}
\end{center}
\caption{\label{F-S1t0}Left:  Time difference between the acoustically determined time of bubble formation and associated PMT trigger, as a function of the bubble position.  The slope of the observed correlation matches the speed of sound in xenon to 20\% \cite{REFPROP}.  Of the 147 bubble events with coincident scintillation pulses, an expected 0.2 are accidental coincidences.  
Right: Scintillation spectrum from a ${}^{57}$Co 122-keV gamma source.  No change in the spectrum is observed between the compressed and superheated states.  The width of the peak is due to the large spatial variation in light-collection efficiency in the chamber.
}
\end{figure}

Figure~\ref{F-event} shows the data streams recorded for each pressure cycle.
These include the pressure and temperature history for the expansion, a sequence of images before and after the bubble trigger, an acoustic record for the event digitized at 2.5~MHz, and a waveform for each xenon PMT trigger throughout the expansion, digitized at 1~GHz by a Keysight U5309A digitizer. The Keysight digitizer operates in a ``triggered simultaneous acquisition and readout'' mode for zero dead time in the PMT data stream, and 
the discriminator used to trigger waveform acquisition
has an estimated 40\% efficiency for single photoelectrons in the November 2016 data.  The discriminator output is suppressed while the LEDs for camera illumination are on to avoid digitizing the $\sim$10-kHz single-photoelectron rate generated by the LED illumination.  The LED gate and a xenon-muon coincidence logic signal are digitized with the acoustic waveform.
Bubbles are correlated with specific scintillation pulses by using the acoustic signal to identify the time of bubble formation, as seen in Fig.~\ref{F-event}.  The distribution of lag times between the scintillation and acoustic signals is shown in Fig.~\ref{F-S1t0}.

We report on 36 live-hours of exposure taken in November 2016, including  background data and exposures to 1-$\mu$Ci~$^{252}$Cf and 175-$\mu$Ci~$^{57}$Co  sources for neutron and gamma calibrations, respectively.  All background and $^{252}$Cf data are taken at $-60^\circ$~C (8.3-keV base threshold), and $^{57}$Co data are taken at both $-60^\circ$~C and $-55^\circ$~C (4.2-keV base threshold).  Thresholds below 4.2~keV were inaccessible due to boiling near the top of the thin annular region between the vials.  This boiling is likely the result of hot spots produced by blackbody radiation, and efforts to reduce the blackbody load from the camera view port have succeeded in lowering the achievable thermodynamic threshold from an initial limit of 30~keV in June 2016 to the 4.2-keV value reported here.  There is no indication that we have reached a fundamental limit to our ability to superheat xenon, and work to improve thermal control and lower the achievable threshold continues.  The additional radiation shielding added for the November run obscured one of the two chamber images, sacrificing 3D position reconstruction for the November data.  Fortunately, no nucleation on the walls of the vessel outside the annular region is observed, so no position-based cuts are necessary.

The $^{57}$Co 122-keV gamma-ray source is used both to calibrate the scintillation response of the chamber and to look for bubble nucleation by gamma interactions in superheated xenon.  The scintillation response of the chamber is measured with the source 74~cm from the target volume for a 335-Hz interaction rate, while for bubble nucleation tests the source is placed immediately outside the cryostat wall, giving a 24.7-kHz interaction rate.  Scintillation spectra are taken at both $-55^\circ$~C and $-60^\circ$~C, from 30~to 200~psi absolute, as shown in Fig.~\ref{F-S1t0}.  No dependence on pressure or temperature is seen, as expected given the small (2\%) density change over this range and the limited resolution of the detector.  The spectrum peaks at 30 photoelectrons, indicating a total photon detection efficiency of 0.4\%.  Our light-collection model translates this to 0.5\% on average for a uniform source, with a strong $z$~dependence (up to a factor of 3) in light-collection efficiency.  The average photon detection efficiency corresponds to an expectation of one photoelectron for a 21-keV nuclear recoil \cite{LUX-DD}.

The high-rate $^{57}$Co data include four single-bubble events in 516~s at the 4.2-keV thermodynamic threshold.  We cannot match bubbles to scintillation pulses in the high-rate data, so we cannot say whether these bubbles are coincident with 122-keV photoabsorption events.  The observed rate is slightly higher than the average background rate of 1.6~mHz, but, given the observed non-Poisson variations in the background rate, we do not take this as evidence for bubble nucleation by gamma rays.  Without background subtraction, we place a  90\% C.L. upper limit of $6.3\times10^{-7}$ on the bubble nucleation efficiency for gamma rays in xenon at $E_T$=4.2~keV.  This same gamma sensitivity is measured in CF$_3$I at $E_T$=5.5~keV, and the extrapolated sensitivity in CF$_3$I at 4.2~keV is $6\times10^{-6}$ \cite{AlansThesis}, an order of magnitude higher than the limit in xenon.  This supports the hypothesis that bubble nucleation by gamma rays in xenon is suppressed by the production of scintillation light.

\begin{figure}
\begin{center}
\includegraphics[width=250pt, clip=true]{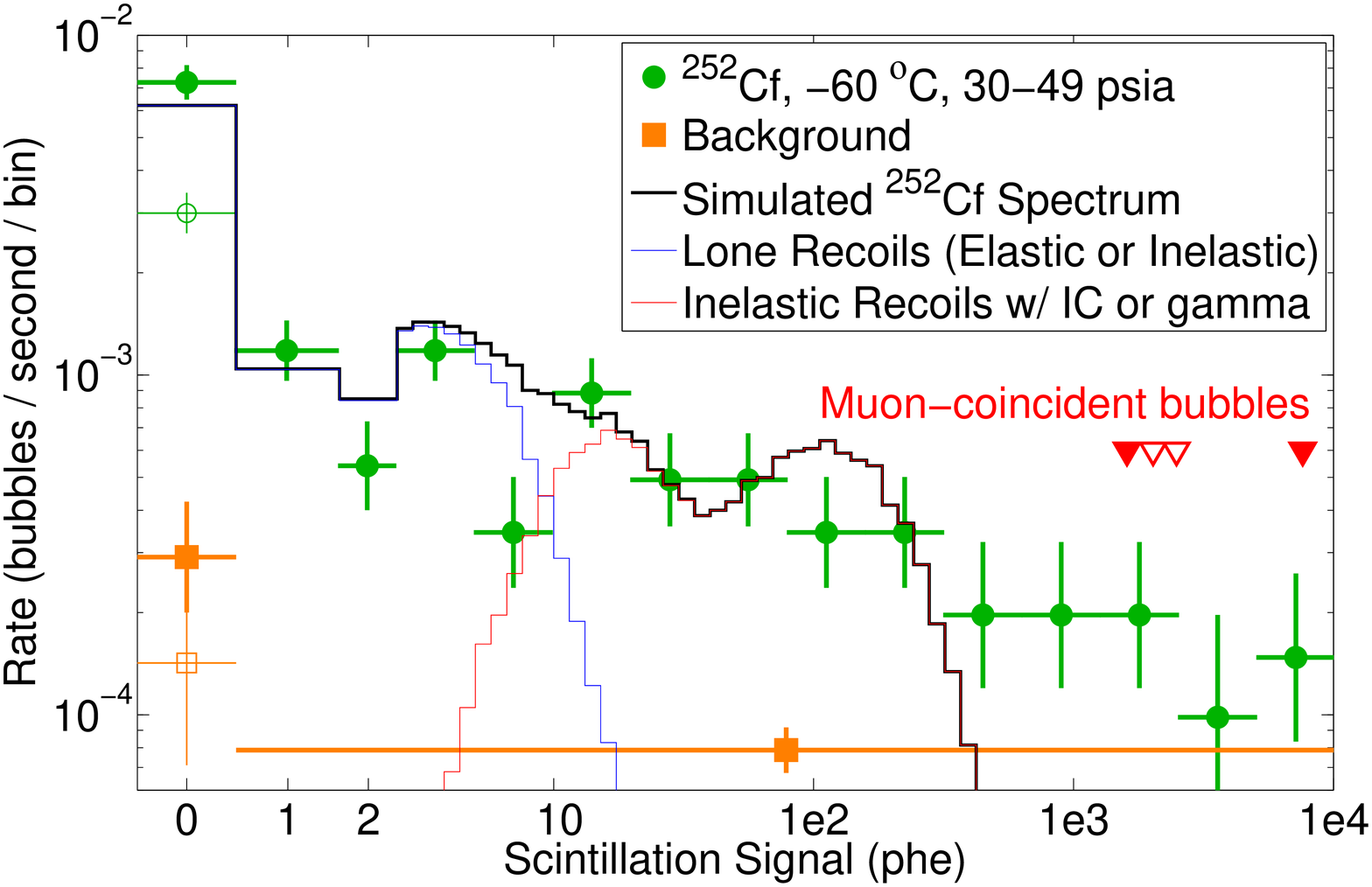}
\end{center}
\caption{\label{F-fullspectrum}Spectrum of scintillation pulses accompanying bubbles in the $^{252}$Cf (green circles) and background (orange squares) data.  The first three bins correspond to 0-, 1-, and 2-photoelectron pulses, where 0 indicates no PMT trigger.  Subsequent bins each span a factor of 2.  Digitized PMT data are unavailable for much of the background exposure, so the background is shown in two bins corresponding to 0 and $\ge$1 photoelectrons.  The data cover thermodynamic thresholds from 8.3~to 15.0~keV.  No significant threshold dependence is seen in the rate of $\ge$1-photoelectron events in either data set.  The rate of 0-photoelectron events is divided into exposures at 8.3--8.6~keV (solid point) and 8.6--15.0~keV (empty point) for both $^{252}$Cf and background data. The four red triangles indicate tagged muon-coincident bubbles, including two in the $^{252}$Cf exposure (solid triangles) and two in the background data (empty triangles).  Also shown is a simulated scintillation spectrum for $^{252}$Cf, selecting only events with nuclear recoils $>$15~keV.  The simulation is divided into lone recoils (elastic scatters or inelastic scatters with escaping gammas) and inelastic scatters with scintillation generated by associated electron recoils.}
\end{figure}

We observe bubbles with coincident scintillation in both the background and $^{252}$Cf data sets, as shown in Fig.~\ref{F-fullspectrum}.  From low- to high-scintillation yield, these bubbles are nucleated by elastic neutron scattering (scintillation produced only by the nuclear recoil), inelastic neutron scattering (scintillation primarily from internal conversion electrons or gamma interactions following the inelastic collision), and cosmic muons (scintillation primarily from the muon track with the bubble produced by a single muon-nucleus elastic scatter, similar to the pion-nucleus scattering observed in Ref.~\cite{CIRTE}).
Four bubbles were coincident with both a xenon scintillation signal and a hit in the scintillator paddle above the chamber, confirming that cosmic muons are the source of these extremely bright events. Bubbles from alpha decays may also be present in the data, with expected scintillation yields between $10^3$ and $2\times10^3$ photoelectrons.

The zero-photoelectron bin in Fig.~\ref{F-fullspectrum} indicates bubbles for which the PMT trigger was active (i.e., not during the camera exposure gate) but no PMT trigger was received.  The rate in this bin shows a strong dependence on $E_T$, consistent with the interpretation that these events are low-energy nuclear recoils.


The  3.1-h $^{252}$Cf exposure at the base threshold $E_T$=8.3~keV contains 160 single-bubble and two double-bubble events.  This is consistent at 1$\sigma$ with the absolute rate predicted by a Monte Carlo simulation of our system using the {\sc mcnpx-Polimi} software package \cite{MCNP} for nuclear recoil bubble nucleation thresholds of $19\pm6$~keV, where the range is dominated by the 30\% uncertainty in our source strength.  The observed multiplicity ratio is consistent with a nuclear recoil energy threshold $\ge$11~keV.

Figure~\ref{F-fullspectrum} also shows a simulated scintillation spectrum derived from the {\sc mcnpx-Polimi} output after applying a 15-keV nuclear recoil threshold.  The postprocessing to produce this spectrum adds electron recoils following inelastic collisions, generates  scintillation light from electron and nuclear recoils according to the best-fit Lindhard model presented in Ref.~\cite{LUX-DD} as implemented in Ref.~\cite{NEST}, and propagates scintillation photons through an optical geometry tuned to fit the observed $^{57}$Co spectrum.   Systematic uncertainties in the source strength, efficiency for triggering on single photoelectrons, and absorption of scintillation light at the walls of the chamber limit our ability to further constrain the nuclear recoil threshold, but the simulated spectrum is qualitatively consistent with observations.  This supports the claim that bubble nucleation by nuclear recoils is not significantly suppressed by scintillation light, nor is scintillation production strongly affected by bubble nucleation.  Future neutron calibrations using $^9$Be$(\gamma,n)$ sources \cite{GammaN} will precisely determine the low-threshold sensitivity of this technique.


\begin{acknowledgements}
We acknowledge all members of the PICO Collaboration.  This work would not have been possible without the technical and scientific developments made by PICO and its predecessors.  We particularly thank Ilan Levine, Edward Behnke, Haley Borsodi, and Thomas Nania at Indiana University South Bend for producing the piezoelectric acoustic transducers used in this work, with the support of National Science Foundation (NSF) Grant No. 1506377.  We also acknowledge technical support from the staff of the Northwestern University Research Shop, as well as support  from  Fermi  National
Accelerator  Laboratory  under  Contract  No.   DE-AC02-07CH11359, and Pacific Northwest National Laboratory,
which is operated by Battelle for the U.S. Department of Energy under Contract No.  DE-AC05-76RL01830.  Thanks go to Tom Shutt and Dan Akerib of SLAC for providing the cryostat used in this work.
This material is based upon work supported by the U.S. Department of Energy (DOE) Office of Science, Office of High Energy Physics under Award No. DE-SC-0012161 and by a DOE Office of Science Graduate Student Research (SCGSR) award.
\end{acknowledgements}

\end{document}